\documentclass{iau}
\usepackage{graphicx,url}

\title[IAUS291.~~New timing solutions for RRATs] 
{New timing solutions for RRATs}

\author[B. Cui et al.]   
{Bingyi Cui$^1$, Jason Boyles$^{1,2}$, Maura McLaughlin$^1$ \\
\and Nipuni Palliyaguru$^1$}

\affiliation{$^1$Dept. of Physics, West Virginia University,  Morgantown, WV 26506, United States \\ email: {\tt bcui@mix.wvu.edu} \\[\affilskip]
$^2$Dept. of Physics and Astronomy, Western Kentucky University,  Bowling Green, KY 42101, United States}

\pubyear{2012}
\volume{291}  
\jname{\mbox{Neutron Stars and Pulsars: Challenges and Opportunities after 80 years}}
\editors{J. van Leeuwen, ed.} 
\begin{document}

\maketitle

\begin{abstract}
The rotating radio transients are sporadic pulsars which are difficult
to detect through periodicity searches. By using a single-pulse search
method, we can discover these sources, measure their periods, and
determine timing solutions. Here we introduce our results on six RRATs
based on Parkes and Green Bank Telescope (GBT) observations, along with a comparison of the spin-down properties of RRATs and normal pulsars.
\keywords{Pulsars: individual, stars: kinematics, stars: rotation}
\end{abstract}

\firstsection 

\section{Introduction}

Rotating Radio Transients (RRATs) are pulsars from which we detect
only sporadic radio bursts, making them difficult to detect in
traditional periodicity searches (McLaughlin et al. 2006). They were
first discovered through  single-pulse search  reprocessing of the
Parkes Multibeam Pulsar Survey data. Currently $\sim$70 of these sporadic
pulsars are known. 
For these, we can calculate
times-of-arrival and determine timing
solutions, by using single pulses instead of the commonly used folded
  profiles.
Timing solutions are crucial to understand their relation 
 to other pulsars and the nature of their
emission.  

Here we introduce our results for six RRATs, along with a comparison between RRATs and normal pulsars. Five of these RRATs were discovered in a re-analysis of the Parkes Multibeam Survey data and one was discovered through a GBT drift-scan survey.

\section{Single-pulse search}

\begin{figure}[tb]
	\begin{center}
  	\includegraphics[trim=220 0 0 0, clip,width=4.9cm,angle=-90]{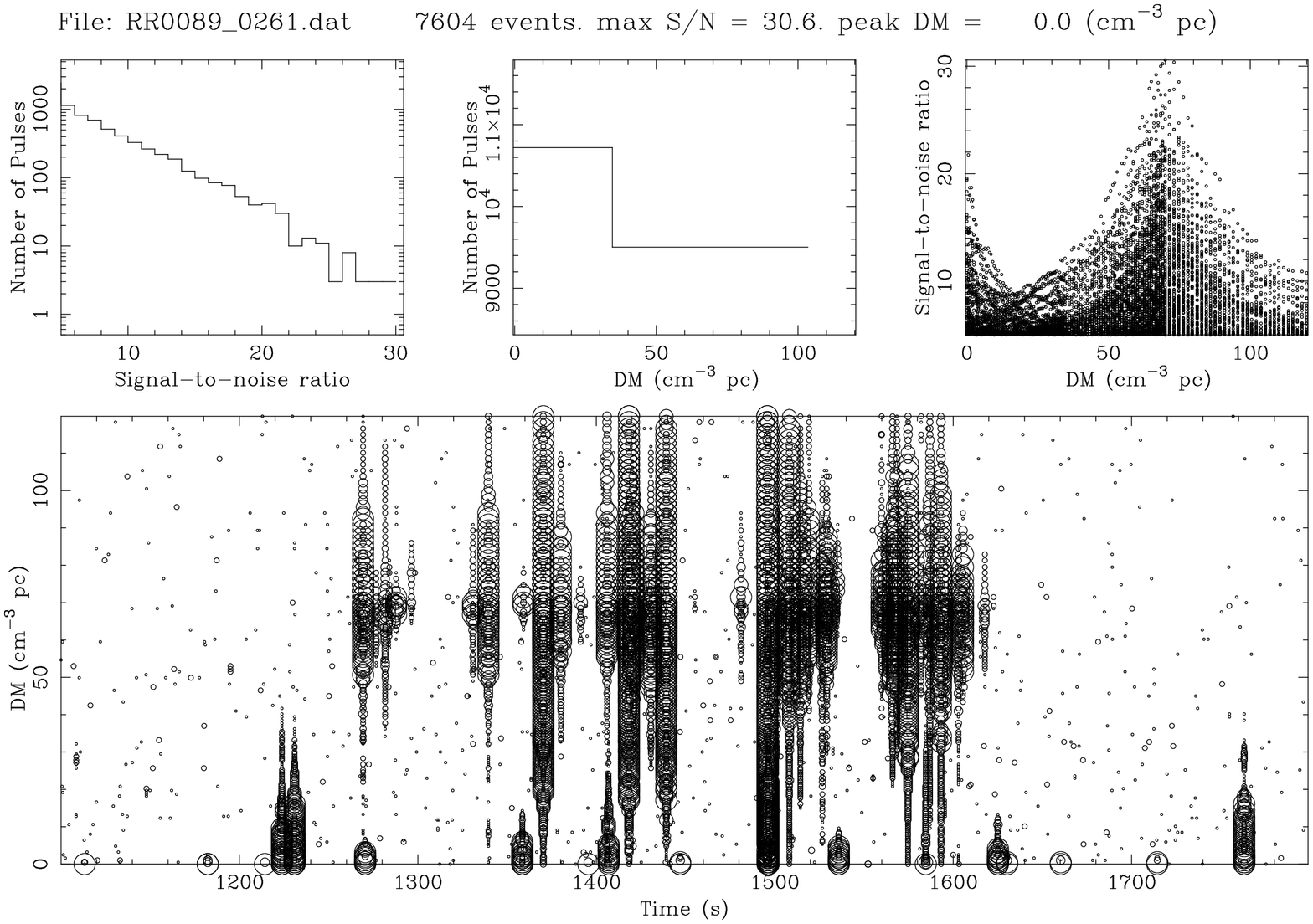}%
\vspace*{-1mm}%
  	\caption{Single-pulse search plot for PSR J1048$-$5838. This
          plot shows a 16-minute portion of a 60-minute Parkes
          1.4\,GHz observation during which the RRAT is `on'. Pulses
          are plotted at the detection DM and time, with size
          proportional to pulse signal-to-noise.}
 	 \label{fig1}
  	\end{center}
\end{figure}

Classical search algorithms based on Fourier techniques or folding
do not detect RRATs. In stead we search for individual pulses with
signal-to-noise ratio above some threshold (typically 5$\sigma$) in a
number of trial-DM time series. Then, once a RRAT is discovered, the
first step in our timing analysis is to identify which pulses are from
the RRAT. We do this by searching for pulses which are brighter at the
DM of the RRAT than at zero DM. Figure \ref{fig1} shows an example of
the single-pulse search output for a nearly one-hour observation of
PSR J1048$-$5838. We see very bright bursts peaking at DM of 69
pc/cm$^3$ which differ from the signals peaking at DM of 0 pc/cm$^3$
that are due to radio frequency interference (RFI). The pulsar only turns `on' for six minutes of this observation. 

\section{Timing solutions}

To get a timing solution for a RRAT, we must first calculate the spin period. We do this by measuring differences between pulse arrival times and calculating the greatest common denominator. Once a period is known, we bin the data into single pulse periods and calculate times-of-arrival (TOAs) as the peak for each detected pulse
.
We have calculated timing solutions for six RRATs. PSR J1048$-$5838
has the longest span of observation of these RRATs: four years of
post-discovery timing observations and a 13-year span including the
discovery. Note that this RRAT was found in an even later reprocessing
of the Parkes Multibeam Survey after the initial RRATs (Keane
2010). The timing residuals shows good phase connection to its earlier
discovery. We have calculated timing solution for this and five other
RRATs. The periods are: 1.231\,s (PSR J1048$-$5838), 0.503\,s (PSR
J1623$-$0841), 1.818\,s (PSR J1739$-$2521), 1.320\,s (PSR J1754$-$3014),
0.933\,s (PSR J1839$-$0141), and 0.414\,s (PSR J1848$-$1243). They have
surface magnetic fields ranging from 4$\times$10$^{11}$\,G to
4$\times$10$^{12}$\,G, spin-down luminosities ranging from
2$\times$10$^{30} $ ergs s$^{-1}$ to 6$\times$10$^{32} $ ergs
s$^{-1}$, and characteristic ages ranging from 1.6 Myr to 15
Myr.\footnote{Latest RRAT solutions can be found in RRATalog:
  \\\url{www.as.wvu.edu/~pulsar/rratalog}}

\section{Pulse profiles and long timescale periodicities}

\begin{figure}[!htb]
\centering
 \begin{minipage}[b]{0.3\textwidth}%
\centering \includegraphics[width=2.4cm]{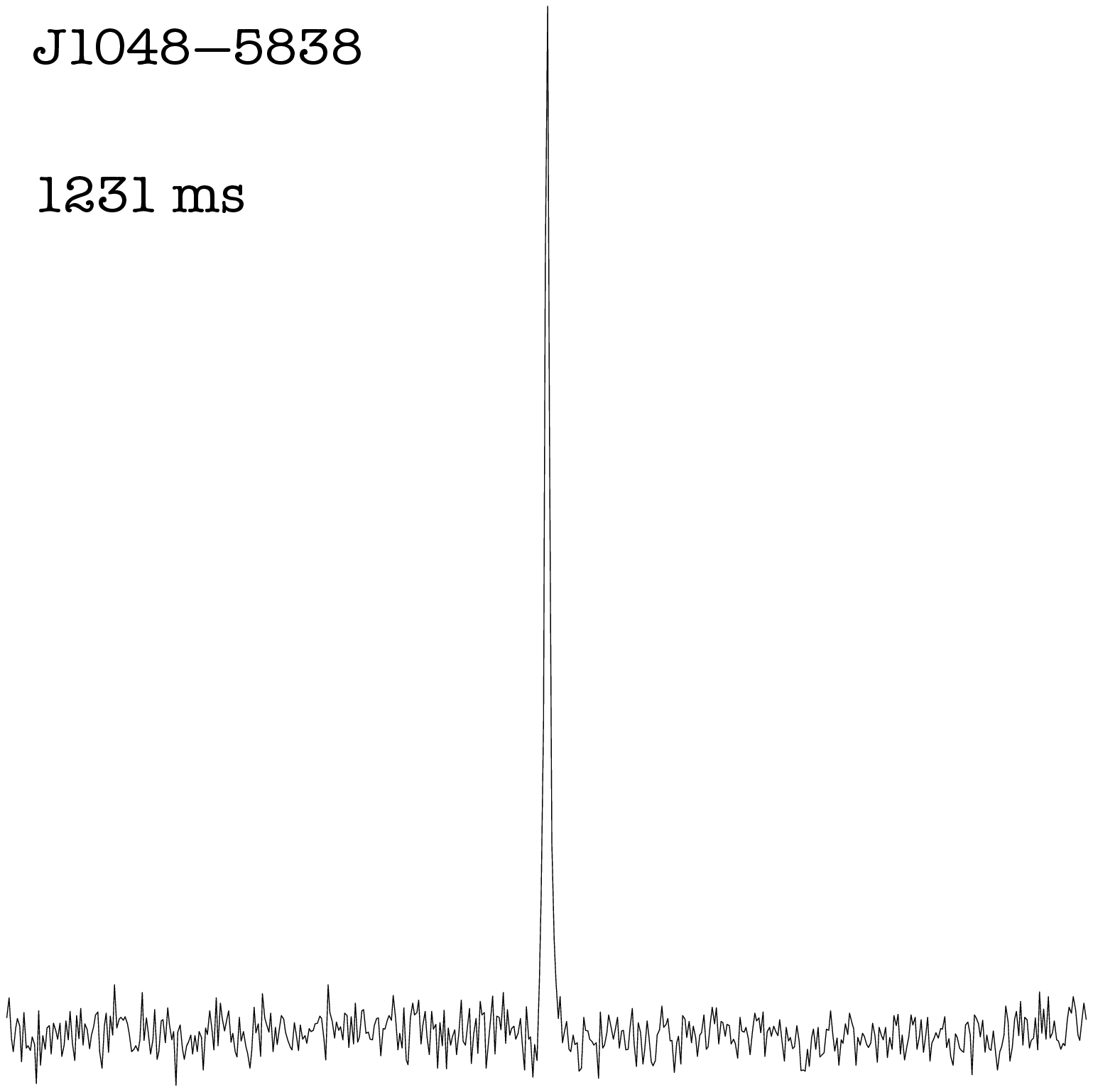}
\vspace*{0.2cm}
\end{minipage}%
 \begin{minipage}[b]{0.3\textwidth}%
\centering \includegraphics[width=2.4cm]{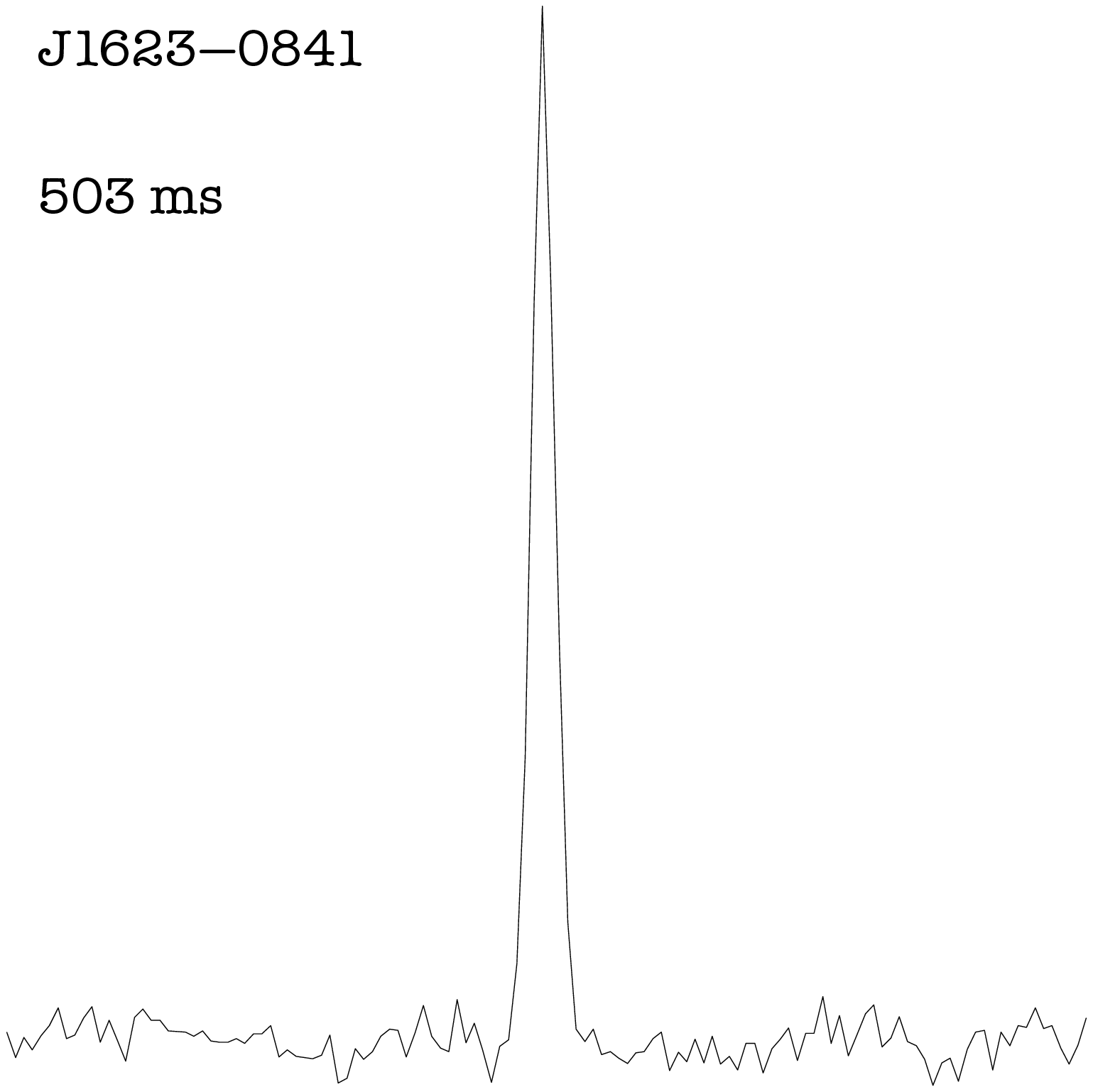}
\vspace*{0.2cm}
 \end{minipage}%
 \begin{minipage}[b]{0.3\textwidth}%
\centering \includegraphics[width=2.4cm]{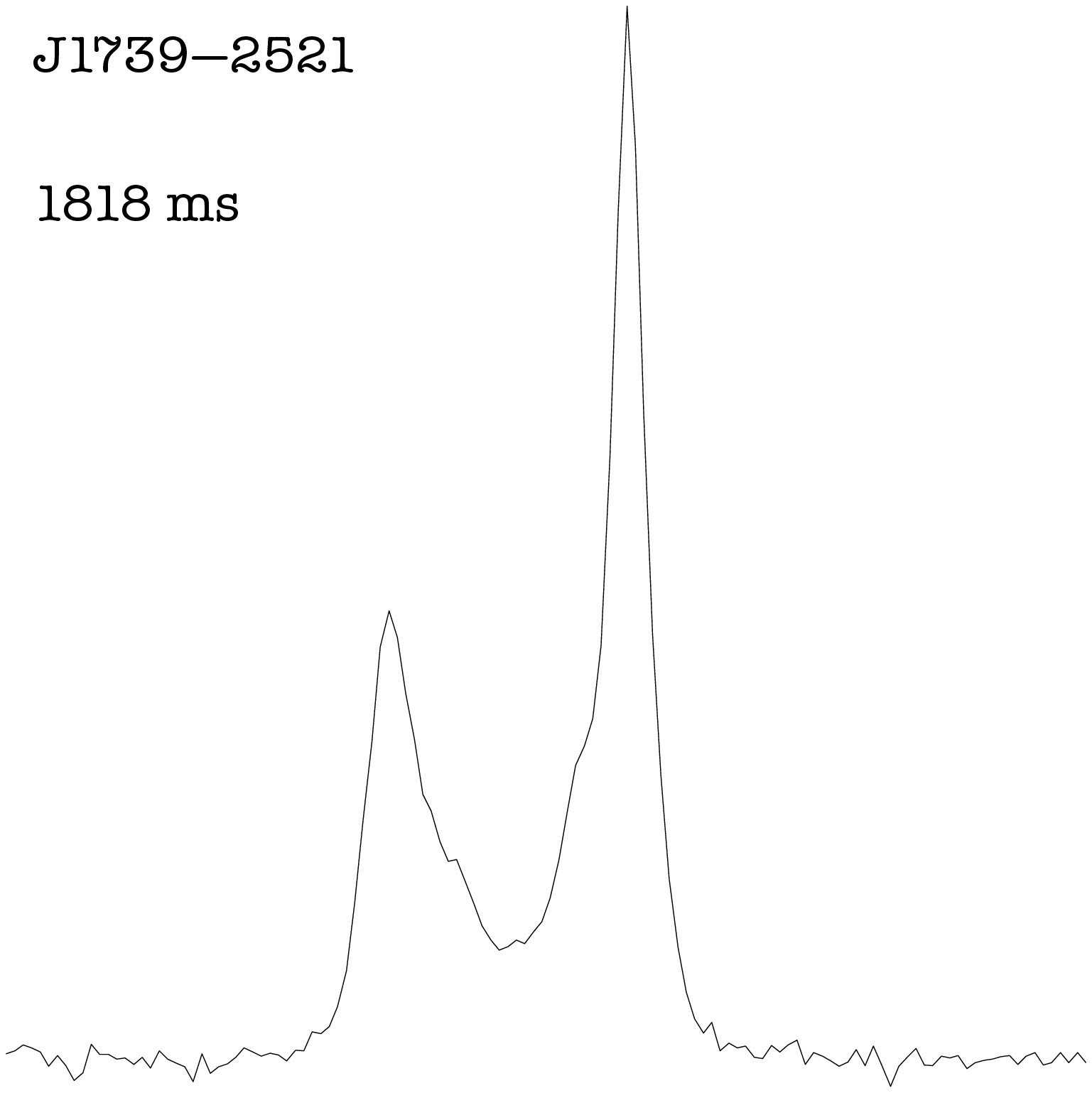}
\vspace*{0.2cm}
 \end{minipage}
\vfill
 \begin{minipage}[b]{0.3\textwidth}%
\centering \includegraphics[width=2.4cm]{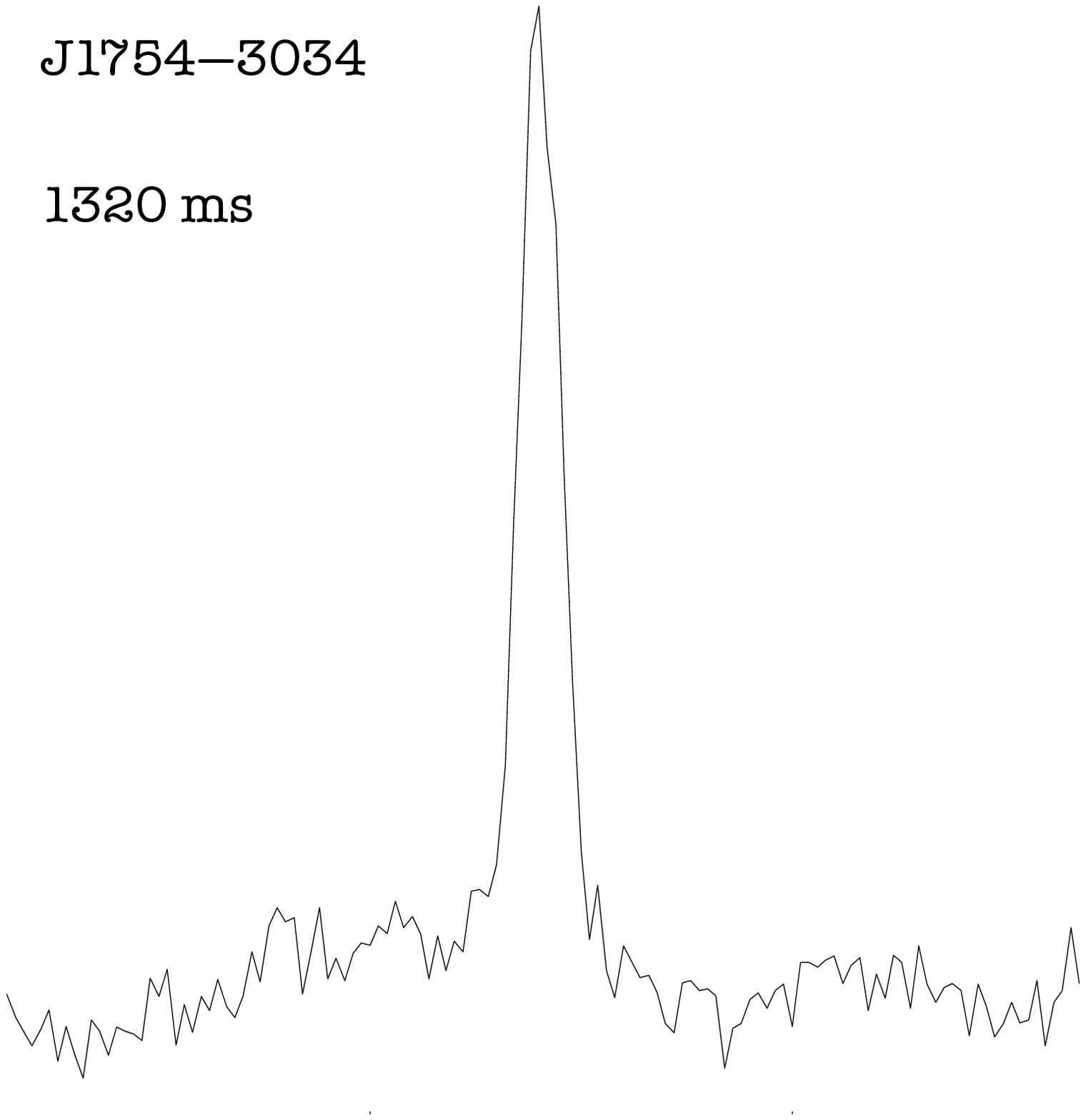}
 \end{minipage}%
 \begin{minipage}[b]{0.3\textwidth}%
\centering \includegraphics[width=2.4cm]{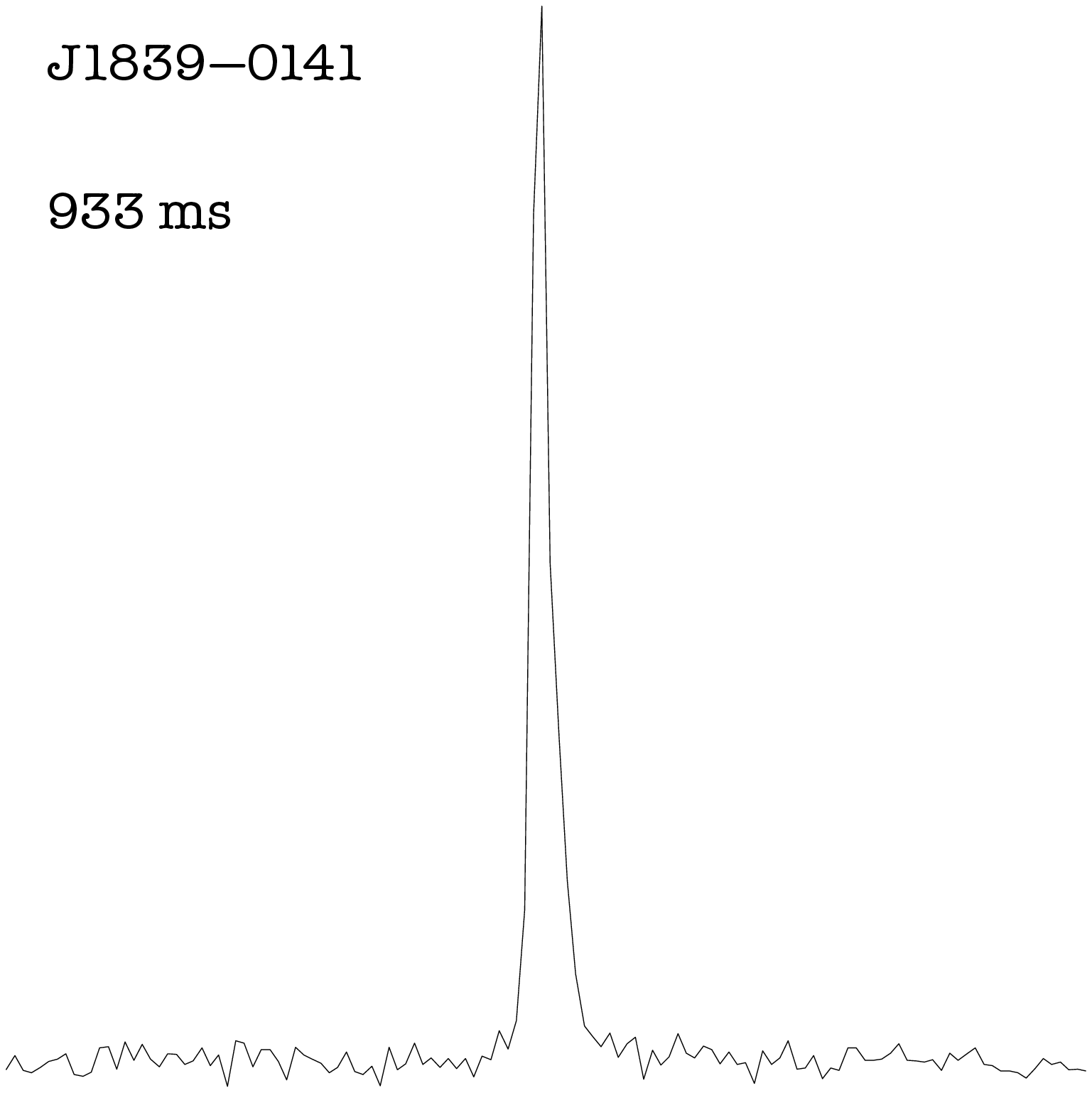}
 \end{minipage}%
 \begin{minipage}[b]{0.3\textwidth}%
\centering \includegraphics[width=2.4cm]{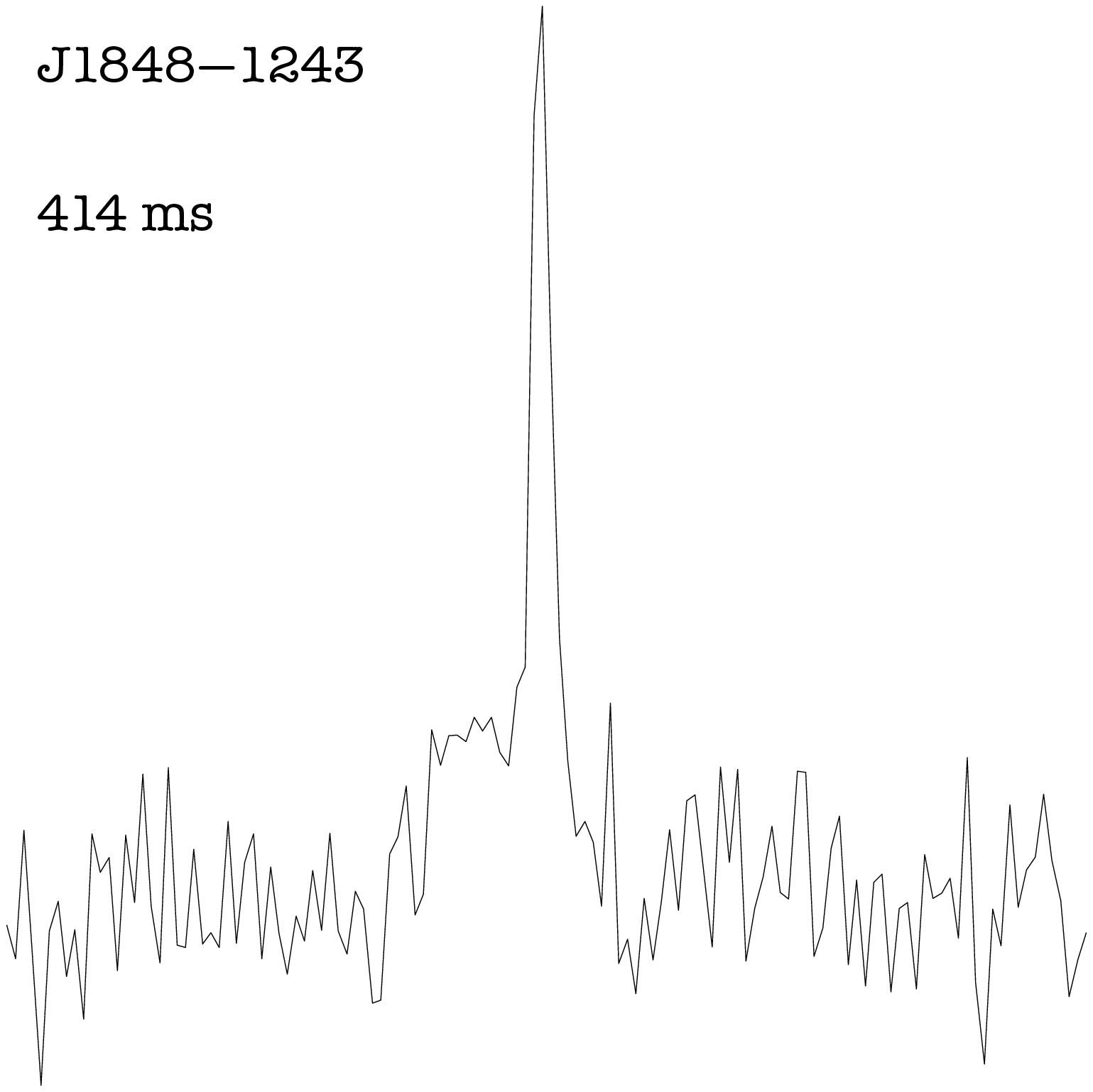}
 \end{minipage}%
\caption{Pulse profiles of 6 RRATs. The profile of PSR J1048$-$5838 is based on 1.4\,GHz observations with Parkes telescope. The others are from 820-MHz observations with the GBT.} 
\label{fig2}
\end{figure}

Profiles of the six RRATs are presented in Figure \ref{fig2}. Here,
pulse profiles of PSRs J1739$-$2521 and J1839$-$0141 are sums of data
during the $\sim$minute-long time periods when the RRATs are `on'. The
profile of PSR J1048$-$5838 is a sum of all detected individual single
pulses. The other profiles are created by folding each observation and
summing all the profiles. Note that these three objects (PSRs
J1623$-$0841, J1754$-$3014 and J1848$-$1243) were detected through
single-pulse searches but are typically detectable in follow-up
observations by folding all of the data. It is clear that the RRATs
are a diverse group of objects with varying properties.

We have applied Lomb-Scargle analysis (Scargle 1982) on the unevenly sampled pulse arrival times of PSR J1048$-$5838 to see if there are any periodicities in the timeseries. The spectrum gives the most significant periodic signal to be of period 19.15 hours with significance level of 1.95 $\sigma$. This is much lower significance than the periodicities of similar timescale reported by Palliyaguru et al. (2011) for other RRATs.

\section{Population of RRATs: Why different?}

\begin{figure}[t]
 \begin{minipage}[b]{0.54\textwidth}%
\centering \includegraphics[trim=350 0 0 0, width=2.05cm, angle=-90]{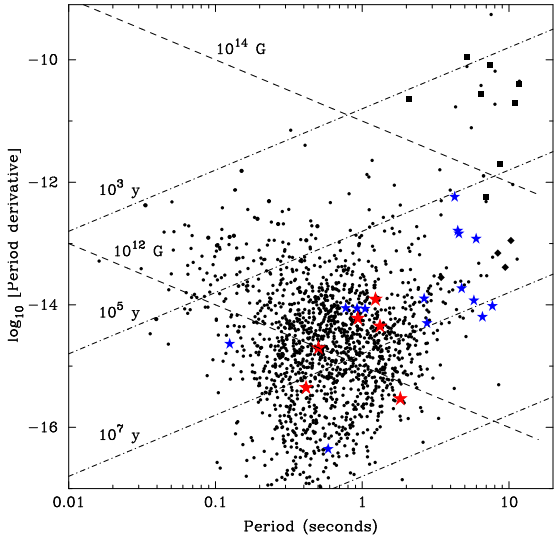}
 \end{minipage}%
\hspace*{5mm}%
\begin{minipage}[b]{0.44\textwidth}%
\caption{$P -\dot{P}$ diagram of solved RRATs and pulsars. New RRATs are shown as red stars and previously timed RRATs as blue stars. The black squares are magnetars, and black diamonds indicate x-ray isolated neutron stars. Lines of constant magnetic field (dashed) and characteristic age (dot-dashed) are shown. The KS test gives 1.12$\times$10$^{-19}$, 2.45$\times$10$^{-4}$, 1.94$\times$10$^{-5}$, 0.16 and 0.04 probabilities that the period, period derivative, magnetic field, and characteristic age, spin-down energy-loss rate, respectively, were derived from the same distribution as those for other pulsars.} 
\label{fig3}
 \end{minipage}
\end{figure}

At this time, 21 of roughly 70 RRATs have timing solutions with period
and period derivative, shown on the $P-\dot{P}$ diagram in Figure
\ref{fig3}. We applied the Kolmogorov-Smirnov test (KS test) to the
RRAT and normal pulsar populations to see how their spin-down
properties compare (see Figure \ref{fig3}).  The largest differences between the two groups are
found in the distributions of period and magnetic field. While selection effects may be responsible
for some of the period dependence, as longer period pulsars are more
likely to be detected with higher signal-to-noise ratio in
single-pulse searches (McLaughlin \& Cordes 2003), the difference in
period derivative distributions hints that there is a fundamental
difference in these populations.

\begin{thebibliography}{}

\bibitem[Keane \etal\ (2010)]{Keane10}
{ Keane, E.F., Ludovici, D.A., Eatough, R.P., \etal} 2010,
\textit{MNRAS}, 1057, 1068

\bibitem[Lorimer \& Kramer (2005)]{LorimerKramer05}
{ Lorimer D.R., \& Kramer M.} 2005,
\textit{Handbook of Pulsar Astronomy} (Cambridge University Press),

\bibitem[McLaughlin \& Cordes (2003)]{CordesMcLaughlin03}
{ McLaughlin, M.A., \& Cordes, J.M.} 2003,
\textit{ApJ}, 982, 996

\bibitem[McLaughlin \etal\ (2006)]{McLaughlin06}
{McLaughlin, M.A., Lyne, A.G., Lorimer, \etal} 2006,
\textit{Nature}, 817, 820

\bibitem[Palliyaguru \etal\ (2011)]{Palliyaguru11}
{ Palliyaguru, N.T., McLaughlin, M.A., Keane, E.F., \etal} 2011,
\textit{MNRAS}, 1871, 1880

\end{thebibliography}
\end{document}